\documentstyle[amsmath,amssymb,aps,epsf,eqsecnum,graphicx]{revtex}

\begin{document}
\draft
\title{Induced measures in the space of mixed quantum states}
\author{Karol
{\.Z}yczkowski$^{1}$\footnote{On leave from Instytut Fizyki,
Uniwersytet Jagiello{\'n}ski, ul. Reymonta 4, 30-059 Krak{\'o}w, Poland}
and Hans-J{\"u}rgen Sommers$^2$
}
\address{$^1$Centrum Fizyki Teoretycznej, Polska Akademia Nauk, \\
Al. Lotnik{\'o}w 32/44, 02-668 Warszawa, Poland}
\address{$^2$ Fachbereich Physik,
Universit\"{a}t-Gesamthochschule Essen\\
 Essen 45117, Germany}
\date{\today}
\maketitle

\begin{abstract}
We analyze several product measures in the space of mixed quantum
states. In
particular we study measures induced by the operation of partial tracing.
The natural, rotationally invariant measure on the set of all pure states
of a $N \times K$ composite system, induces a unique measure in the
space of
$N \times N$ mixed states (or in the space of $K\times K$ mixed states,
if the reduction takes place with respect to the first subsystem).
For $K=N$ the induced measure is equal to the Hilbert-Schmidt
measure, which is shown to coincide with the measure
induced by singular values of non-Hermitian random
Gaussian matrices pertaining to the Ginibre ensemble.
We compute several averages with respect to this measure and show
that the mean entanglement of $N \times N$ pure states
behaves as ln$N-1/2$.
\end{abstract}


\medskip
\begin{center}
{\small e-mail: karol@cft.edu.pl \ \quad \
sommers@next30.theo-phys.uni-essen.de}
\end{center}


\section{Introduction}

Recent developments in the emerging field of quantum information
increased an interest in investigation of the properties of the set of
density matrices of a finite size. To characterize quantitatively
properties of typical density matrices it is necessary to define a
certain measure $\mu$ in this set, which will enable one to compute the
desired averages over all $N \times N$ density matrices  with respect
to this measure.  Only having chosen the measure $\mu$ it makes sense to
ask about the probability to find a random density matrix representing
an entangled state
or to compute the average entanglement of mixed quantum states
\cite{ZHSL98,Sl99a,Zy99}. Alternatively one may search for a statistical
ensemble of density matrices which corresponds to minimal prior
knowledge about a quantum system \cite{Ha98,Sl98}.

In the space of pure states of an $N$ dimensional
Hilbert space ${\cal H}_N$, isomorphic with the complex projective space
${\mathbb C}P^{N-1}$,
there exist a unique 'natural' measure, $P_N(|\phi\rangle)$,
induced by the Haar measure on the unitary group $U(N)$.
In other words, the random pure state, defined by action of a random
unitary matrix on an arbitrary reference state,
$|\phi\rangle=U|\phi_0\rangle$,
may be represented (in an arbitrary basis) as a certain row
(or column) of the random matrix $U$.
The measure $P_N(|\phi\rangle)=P_N(U|\phi\rangle)$ in the space of pure
states is thus
distinguished by the rotational invariance
of the Haar measure \cite{Wo90,Jo90,Jo91}.

This property of the set of pure states may suggest that there exists a
natural measure in the space of mixed quantum states ${\cal M}_N$,
distinguished in an analogous way. However, this might not be the case:
it is not at all simple to select a single measure in ${\cal
M}_N$ and provide rational arguments for
its superiority  \cite{Wo90,Ha98,Sl99a,Zy99}.
Any density matrix $\varrho$ is Hermitian and may be
diagonalised by an unitary matrix $U$.
It is thus tempting to assume that
the distributions of eigenvalues and eigenvectors of
$\varrho$ are independent, so $\mu$
factorizes into a {\sl product measure}  $\Delta \times \nu$.
For the measure $\nu$ on the unitary group, which determines the
statistics of the eigenvectors forming $U$, one may
take the unique Haar measure on $U(N)$. On the other hand,
it is much more difficult to find convincing arguments,
which allow one to pick out
the unique measure $\Delta$ defined on the simplex
of eigenvalues.

To consider a possibility of defining induced measures in
the $(N^2-1)$ dimensional space
${\cal M}_N$ it is worth to discuss the procedure of {\sl purification}.
Any mixed state $\varrho$, acting on ${\cal H}_N$,
may be purified by finding a pure state $|\Psi\rangle$
in the composite Hilbert space
${\cal H}_N \otimes {\cal H}_N$,
such that $\varrho$ is given by the partial tracing over the auxiliary
subsystem,
\begin{equation}
|\Psi\rangle \longrightarrow {\varrho}
={\rm Tr}_2\Bigl(
|\Psi\rangle \langle \Psi | \Bigr)\ .
\label{puriff}
\end{equation}
In a loose sense the purification
corresponds to treating
any density matrix of size $N$
as a vector of size $N^2$.
For a given $\varrho$ finding a corresponding
 pure state $|\Psi\rangle$ does
not have a unique solution since
\begin{equation}
\varrho = {\rm Tr}_2 \Bigl(
|\Psi\rangle\langle\Psi|\Bigr) =
{\rm Tr}_2\Bigl(
|\Psi'\rangle \langle \Psi' | \Bigr)
\quad \quad \quad {\rm where}
\quad \quad \quad  |\Psi'\rangle = ({\mathbb I}\otimes U_2 )| \Psi \rangle
\label{purif3}
\end{equation}
where $U_2$ is an arbitrary unitary matrix of size $N$
acting locally on the auxiliary subsystem.

It is now simple to formulate the idea of the induced measure:
we choose the natural measure $P_{N^2}(|\Psi\rangle)$ in the space of
purified states, and look for the measure $P_{N,N}(\varrho)$
induced by the operation of partial tracing (\ref{puriff}).
This idea, put forward by Braunstein \cite{Br96},
was further developed by Hall \cite{Ha98}.
He considered the ensemble of pure states in the $NK$ dimensional
space ${\cal H}_N \otimes {\cal H}_K$ distributed according to the
the natural measure $P_{NK}(|\Psi\rangle)$.
The partial tracing over the variables of ${\cal H}_K$
defines uniquely the induced measure $P_{N,K}(\varrho)$
in the space of mixed states ${\cal M}_N$.
Hall found explicitly the probability distributions
$P_{2,K}(\varrho)$ and computed mean entropy averaged over these
measures. The aim of this work is to extend his results providing the
general formula for $P_{N,K}(\varrho)$. Moreover, we
propose other methods of defining induced measures
via the Hilbert-Schmidt space and establish a link between them.

The space of linear operators acting on ${\cal H}_N$,
equipped with the scalar product
$\langle A|B\rangle=$tr$A^{\dagger}B$, is called the Hilbert-Schmidt
space, ${\cal H}_{HS}$. It has $N^2$ complex dimensions and may be
represented by the algebra of complex matrices of size $N$.
Any non-zero matrix $A$ may be projected into the space of
density matrices by \cite{Be98}
\begin{equation}
  A \longrightarrow {\varrho} =
\frac{AA^{\dagger}}{\mbox{tr}AA^{\dagger}} \ .
\label{projHS}
\end{equation}
It is easy to see that $AA^{\dagger}$ is Hermitian
and positive  (strictly speaking, non-negative),
 while the rescaling by
the norm  assures the trace condition tr$\varrho$=1.
Obviously one may consider an alternative, entirely
equivalent projection
 $A \to \varrho'={A^{\dagger}A}/ {\mbox{tr}A^{\dagger}A}$,
 which agrees with (\ref{projHS}) for normal $A$.

Two matrices $A$ and $A'$ produce the same state $\varrho$,
if there exist an unitary $V$, such that $A'=AV$.
An arbitrary measure $P(A)$ in the Hilbert-Schmidt space
induces thus a certain measure in the space $\cal M$ of
density matrices.
The eigenvalues of $\varrho$ are equal to the rescaled squared singular values
of matrix $A$, where $SV(A)=\sqrt{{\rm eig} (A A^{\dagger}) }$
\cite{Horn}. Note that the singular values of $A$ and $A'$
are the same.
For $P(A)$ one may take the
measure of Ginibre ensemble of (non-Hermitian)
random Gaussian matrices. In this paper we prove that the measure
induced in this way is equal to $P_{N,N}(\varrho)$, which is also
shown to coincide with the measure related to the Hilbert-Schmidt
metric.

Our paper is organized as follows. In Section II we
write down the natural measure in the space of pure states
and provide a review of different measures
used in the space of the density matrices.
Section III is devoted to measures induced by partial tracing.
In section IV we discuss possibilities of
generalizing our results
and provide certain families of measures in $\cal M$.
The paper is concluded in section V.

\section{Measures in the space of density matrices}

\subsection{Natural  measure on the space of pure states}

Before concentrating on the set ${\cal M}_N$ of the density matrices
of size $N$, let us  discuss the space of pure states.
For concreteness we  start our considerations with the four dimensional
Hilbert space - the simplest case important from the point of view of
quantum entanglement. The set of pure states of a $4$ dimensional Hilbert
space forms a complex  projective
space ${\mathbb C}  P^{3}$, on which the
natural uniform measure exists. To generate random pure states according to
such a measure on this $6$ dimensional manifold, we take a vector of a random
unitary matrix distributed according to the Haar measure on $U(4)$. The
Hurwitz parametrisation \cite{Hu87} gives

\begin{equation}
| \Psi \rangle =
(\cos \vartheta_3, \sin \vartheta_3 \cos \vartheta_2 e^{i
\varphi_3}, \sin \vartheta_3 \sin \vartheta_2 \cos\vartheta_1 e^{i
\varphi_2}, \sin \vartheta_3 \sin\vartheta_2 \sin \vartheta_1 e^{i
\varphi_1} ),
 \label{param3}
\end{equation}
where $\vartheta_k  \in [0,\pi/2],$ and $\varphi_k \in [0, 2\pi)$
for $k=1,2,3 $.

A uniform distribution over almost all of ${\mathbb C} P^{3}$ is
obtained by choosing the uniform distribution of the 'polar' angles; $%
P(\varphi_k)= 1/ 2\pi$. In analogy to the volume element on the sphere,
the 'azimuthal' angles $\vartheta_k$ should be taken in a nonuniform way,
with the probability density \cite{Hu87}

\begin{equation}
P(\vartheta_k)=k \sin(2\vartheta_k) (\sin \vartheta_k)^{2k-2} \quad \quad
{\rm for} \quad \quad \vartheta_k \in [0,\pi/2], \quad \ \ k=1,2,3.
\label{density2}
\end{equation}
In practice it is convenient to use auxiliary independent random variables $%
\xi_k$ distributed uniformly in $[0,1]$ and to set $\vartheta_k={\rm arcsin}%
\bigl(\xi_k^{1/2k} \bigr)$. In order to generalize the
parametrisation (\ref{param3}) for a vector belonging to the
$N$-dimensional
Hilbert space, one uses $(N-1)$ polar angles $\varphi _{k}$, uniformly
distributed in $[0,2\pi )$, and $(N-1)$ independent azimuthal angles
$\vartheta _{k}$, distributed according to (\ref{density2})
\cite{PZK98}.
The volume element in space ${\mathbb C} P^{N-1}$ reads thus
\begin{equation}
dv = \prod_{k=1}^{N-1} \cos\vartheta_k
(\sin\vartheta_k)^{2k-1} d\vartheta_k d \varphi_k,
\label{volum1}
\end{equation}
while the total volume of the manifold
of pure states
\begin{equation}
V_N = \prod_{k=1}^{N-1} \int_0^{\pi/2} \cos\vartheta_k
(\sin\vartheta_k)^{2k-1} d\vartheta_k
 \prod_{k=1}^{N-1} \int_0^{2\pi} d \varphi_k
=\frac{\pi^{N-1}}{(N-1)!}
\label{volum2}
\end{equation}
 is proportional to the product
of the volume of the $(N-1)$-d torus,
(built of the phases $\varphi_k$),
times the volume of the $(N-1)$-d simplex
${\cal S}_N$, (built of the angles $\vartheta_k$) \cite{Ba00}.
The natural measure (\ref{volum1})
in ${\mathbb C} P^{N-1}$ induces the {\sl uniform}
measure in the simplex ${\cal S}_N$ \cite{Zy99}.
In appendix A we show that the same measure
is given by squared absolute values of
$N$ independent random complex Gaussian numbers $z_i$,
rescaled as $|z_i|^2\to |z_i|^2/\sum_{i=1}^N |z_i|^2$.
Thus a random pure state may be constructed of
$N$ appropriately rescaled complex random numbers
drawn according to the normal distribution.

It is worth to note that the natural measure
(\ref{volum1}) is
compatible with the Fubini-Study metric
on ${\mathbb C} P^{N-1}$, in the sense that
the measure of a certain $\epsilon$-neighbourhood
of any chosen point is the same \cite{Be98}.
The average value of
some function of a state, $X_{\psi}=X(|\psi\rangle)$
(e.g. an observable $\langle X\rangle_{\psi}$),
over the entire
manifold of pure states is then defined as
\begin{equation}
{\bar X} = \frac{1}{V_N} \int_{CP^{N-1}}
 X_{\psi}   dv(|\psi\rangle) .
\label{volum3}
\end{equation}

\subsection{Product measures in the space of mixed states}

Consider a Hermitian, nonnegative matrix $\varrho$ of size $N$,
normalized as tr$\varrho=1$. It represents a mixed state belonging to the
space ${\cal M}_N$ and may be diagonalized by a unitary
rotation. Let $B$ be a diagonal unitary matrix. Since
\begin{equation}
\varrho =U\Lambda U^{\dagger }=UB\Lambda B^{\dagger }U^{\dagger },
\label{diag2}
\end{equation}
the rotation matrix $U$ is determined up to $N$ arbitrary phases
entering $B$. The number of irrelevant phases increases, if the spectrum of $%
\varrho$ is degenerated (see e.g. \cite{ach93}). The total number of
independent variables used to parametrize a density matrix $\varrho$ is equal
to $N^2-1$, provided, no degeneracy occurs. In recent papers discussing the
relative volume of the set of entangled states \cite{ZHSL98,Zy99}, we
considered the product measures
\begin{equation}
\mu =\Delta \times \nu.  \label{muu}
\end{equation}
The first factor defines a measure in the $(N-1)$ dimensional simplex of
eigenvalues entering $\Lambda$, since due to the trace condition
$\sum_{i=1}^{N}\lambda_{i}=1$.
The measure $\nu$, defined in the space of unitary matrices $U(N)$,
 is responsible
for the choice of eigenvectors of $\varrho$. It is natural to take for $\nu$
the Haar measure $\nu_u$ on $U(N)$,
so that the probability distributions are rotationally invariant,
$P(\varrho)=P(U\varrho U^{\dagger})$.
  On the other hand,
as pointed out by Slater \cite{Sl99a}, there exist several possible choices
for the measure $\Delta$.

\subsection{Measures induced by random vectors}

In Ref. \cite{Zy99} we defined a {\sl unitary
product measure} by taking for the
vector ${\vec \Lambda}=\{\lambda_{i}\}$ the squared
moduli of complex elements of a column (say, the first column) of an
auxiliary random unitary matrix $V$ drawn with respect to the Haar measure $%
\nu_{u}$ on $U(N)$
\begin{equation}
\lambda_{i}=|V_{i1}|^{2}.  \label{dii}
\end{equation}
The distribution $P_u({\vec \Lambda})$ obtained in this way
is nothing else but the distribution of components of
a random vector on
 ${\mathbb C} P^{N-1}$,
 uniform on the simplex of eigenvalues \cite{Zy99}.
In a similar way one defines the {\sl orthogonal product measure}
where
$\lambda_{i}=|O_{i1}|^{2}$ and $O$ is an orthogonal matrix drawn with
respect to the Haar measure $\nu_{o}$ on $O(N)$.

Let us stress that the name of the product measure (orthogonal or unitary)
is related to the distribution $\Delta$ on the simplex of eigenvalues, while
the random rotations $U$ are always assumed to be distributed according to
the Haar measure $\nu_u$ in $U(N)$. Thus both measures may be linked
directly to the well known Gaussian unitary (orthogonal) ensembles of random
matrices \cite{Mehta,Haake}, referred to as GUE and GOE. The measure $\mu_o$
is determined by squared components of a real eigenvector of a GOE
matrix, while
the measure $\mu_u$ may defined by components of a complex eigenvector
of GUE matrices.

It is easy to see, that these measures on the simplex are
special
 cases of the Dirichlet distribution \cite{Sl99a}
\begin{equation}
\Delta_s(\lambda_1,\dots,\lambda_{N-1}) = \alpha_s \lambda_1^{s-1}\!\dots\!
\lambda_{N-1}^{s-1}(1-\lambda_1-\!\dots\!-\lambda_{N-1})^{s-1}.
\label{Dirichlet}
\end{equation}
where $s > 0$ is a free parameter
and $\alpha_s$ stands for a normalization
constant. The last component is determined by the trace condition $%
\lambda_N=1-\lambda_1-\dots-\lambda_{N-1}$. The uniform measure $\Delta_1$
describes the unitary measure $\mu_u=\Delta_1\times \nu_u$, while the case $%
s=1/2$ corresponds to the orthogonal measure, $\mu_o=\Delta_{1/2}\times \nu_u
$ \cite{Zy99}. The latter   is related to the Fisher information metric
\cite{Fisher}, the Mahalonobis distance \cite{Ma1}
 and Jeffreys' prior \cite{Jeff}.

It is instructive to consider the limiting cases of the distribution (\ref
{Dirichlet}). For $s \to 0$ one obtains a singular distribution peaked at
the pure states \cite{Sl99a}, while in the opposite limit $s \to \infty$, the
distribution is concentrated on the maximally mixed state $\varrho_*$ described
by the vector ${\vec \Lambda}=\{1/N,\dots,1/N\}$. Changing the
continuous
parameter $s$ one can thus control the average purity of the generated mixed
states.

\bigskip

\subsection{Measures related to metrics}

There exist several different distances in the space of mixed quantum
states - (see e.g. \cite{PS96,ZS00}).
Each metric generates a measure, since one can assume that
drawing random matrices from each ball of a fixed radius is equally
likely. The balls are understood with respect to a given metric.
We shall consider here two most important examples: the Hilbert-Schmidt
metric and the Bures metric.

The Hilbert-Schmidt distance between any two density operators  is given
by the Frobenius (Hilbert-Schmidt) norm of their difference
\begin{equation}
D_{HS}(\varrho_1,\varrho_2)=\sqrt{ {\rm Tr}
 [(\varrho_1 - \varrho_2)^2] }.
  \label{HS1}
\end{equation}
With respect to this metric the set of the two levels mixed states
has the geometry of the Bloch sphere and its interior.
For any dimension $N$ the infinitesimal distance takes a particularly
simple form
\begin{equation}
(ds_{HS})^2=  {\rm Tr}  [(\delta \varrho)^2]
  \label{HS2}
\end{equation}
and allows one to compute the volume element,
normalize it by appropriate constants $C_N$ and to obtain the
probability distribution
\begin{equation}
 P_{HS}(\lambda_1,\dots,\lambda_N) = C_N \delta(1-\sum_{i=1}^N
\lambda_i )  \prod_{j<k}^N (\lambda_j-\lambda_k)^2.
  \label{HS3}
\end{equation}
This joint probability distribution, derived by Hall \cite{Ha98},
defines the {\sl Hilbert-Schmidt measure} $\Delta_{HS}$ in the space of
diagonal matrices, and a
product measure (\ref{muu}) in the space of density matrices.

 Another measure   is related
 with the Bures distance in the space of
mixed quantum states, \cite{bu69,ul76},
\begin{equation}
D_{B}(\varrho_1,\varrho_2)=\sqrt{ 2\bigl(1-{\rm {tr} [(\varrho
_1^{1/2}\varrho_2\varrho_1^{1/2})^{1/2}]\bigr) }},
 \label{Bures1}
\end{equation}
It corresponds to the minimal Fubini-Study distance
between all possible purifications (\ref{puriff})
of both mixed states $\varrho_1$ and $\varrho_2$.
With respect to this distance the $3$-dimensional
set of $N=2$ density matrices exhibits
the geometry of a half of the hypersphere $S^3\in R^4$, with pure states
at the hyper-meridian \cite{Ul92}.

The infinitesimal Bures  distance between $\varrho$ and $\delta \varrho$  was
found by H{\"u}bner
\cite{hu92}
\begin{equation}
(ds_{B})^2=  \frac{1}{2} \sum_{j,k=1}^N \frac
{ |\langle j |\delta \varrho |k\rangle |^2 }{\lambda_j+\lambda_k},
  \label{Bures2}
\end{equation}
where $\lambda_k$ and $|k\rangle$ represent eigenvalues and eigenvectors
of $\varrho$.
Hall computed the corresponding volume element, and received the
 {\sl Bures probability distribution} \cite{Ha98}
\begin{equation}
 P_{B}(\lambda_1,\dots,\lambda_N) = C'_N \frac
 { \delta(1-\sum_{i=1}^N \lambda_i )} {
 (\lambda_1 \cdots \lambda_N)^{1/2} }
 \prod_{j<k}^N \frac{ (\lambda_j-\lambda_k)^2 }{
 \lambda_j+\lambda_k }.
  \label{Bures3}
\end{equation}
This measure was later analyzed by Slater \cite{Sl99b},
who found the normalization constants $C'_N$
for $N=3,4$ and $5$ and together with Byrd  analyzed
the geometry induced by the Bures measure for $N=3$ \cite{BS00}.
The Bures measure is {\sl monotone},
i.e. it does not increase under the action of
completely positive, trace preserving maps \cite{PS96}.

\bigskip

\subsection{$N=2$: Measures in the Bloch ball}

To experience some features of different measures consider the
set of $2 \times 2$ density matrices.
Such a matrix can be expressed by the
Pauli matrices ${\vec \sigma}$ as $\varrho=I/2 +{\vec \sigma}\cdot
{\vec r}$, where ${\vec r}$ is
a three dimensional vector of modulus $r\le 1/2$.
Pure states correspond to the Bloch sphere, $r= 1/2$,
while the center of the ball, $r=0$, corresponds
to the maximally mixed state
proportional to the identity matrix.
 The eigenvalues of $\varrho$ read $\lambda_1=1/2+r$, and
$\lambda_2=1/2-r$.
The rotationally invariant Haar measure $\nu_u$
implies in this case the uniform distribution of the pure states on
the Bloch sphere. The product measures
may differ by a different radial distribution P(r).

The unitary measure $\Delta_1$ is uniform in radius,
$P_u(r)=2$ for $r\in [0,1/2]$,
while the orthogonal measure $\Delta_{1/2}$
defined by (\ref{Dirichlet})  gives the "cosine distribution",
$P_o(r)=4/(\pi\sqrt{1-4r^2})$.
The Hilbert-Schmidt measure (\ref{HS3}) reads for $N=2$
\begin{equation}
P_{HS} (r) = 24 r^2 \quad {\rm for } \quad r\in [0,1/2]
  \label{HS4}
\end{equation}
The quadratic factor relates to the Jacobian in  $3$-D spherical
coordinates, so the  Hilbert-Schmidt measure
corresponds to the uniform coverage of the entire Bloch ball
\cite{Ha98}.

On the other hand, the Bures measure (\ref{Bures3})
implies the following radial distribution, obtained by
substituting $\lambda_1=1/2+r$ and $\lambda_2=1/2-r$
\begin{equation}
P_B(r)=\frac{ 32 r^2}{ \pi \sqrt{1 -4r^2}}.
\label{Bures4}
\end{equation}
where $r\in [0,1/2]$.
As shown in Fig. 1a this distribution is more concentrated
on high purity states with large $r$, located close to the Bloch sphere.

To analyze  the simplest measure $P_{2,2}(\varrho)$ induced by
partial tracing, consider pure state
 $|\Psi \rangle =(x,y,z,t)$ of the  $2\times 2=4$
dimensional Hilbert space ${\cal H}_4$. The
corresponding density matrix reads
\[
\varrho =|\Psi \rangle \langle \Psi |= \left[
\begin{array}{cccc}
xx^{\ast } & xy^{\ast } & xz^{\ast } & xt^{\ast } \\
yx^{\ast } & yy^{\ast } & yz^{\ast } & yt^{\ast } \\
zx^{\ast } & zy^{\ast } & zz^{\ast } & zt^{\ast } \\
tx^{\ast } & ty^{\ast } & tz^{\ast } & tt^{\ast }
\end{array}
\right],
\]
Performing the partial tracing with respect to the second subsystem one gets
the reduced density matrix
\begin{equation}
\varrho _{A}=Tr_{B}(\varrho )= \left[
\begin{array}{cc}
xx^{\ast }+yy^{\ast } & xz^{\ast }+yt^{\ast } \\
zx^{\ast }+ty^{\ast } & zz^{\ast }+tt^{\ast }
\end{array}
\right] =: \left[
\begin{array}{cc}
b & w \\
w^{\ast } & 1-b
\end{array}
\right],  \label{2x2matrix}
\end{equation}
where $b=xx^{\ast }+yy^{\ast }$ and $w=xz^{\ast }+yt^{\ast }$. Its
eigenvalues might be written as $\lambda _{1,2}=\frac{1}{2}\pm r$,
 where $r=\frac{1}{2} \sqrt{(1-2b)^{2}+4|w|^{2}}$
It has been already proved by Hall \cite{Ha98}
that the induced distribution
$P_{2,2}(r)=24r^{2}$ and coincides with the Hilbert-Schmidt
measure  (\ref{HS4}). In the next section we demonstrate
that this relation holds for arbitrary $N$.

 The pure states $|\Psi\rangle \in {\cal H}_4$ may be
represented by the Schmidt decomposition
$|\Psi\rangle= \cos\alpha|00\rangle
+ \sin\alpha |11\rangle $, where $\{
|00\rangle,|01\rangle,|10\rangle,|11\rangle \} $
 denotes the standard basis in ${\cal H}_4$
and $\alpha \in [0,\pi/4] $.
The corresponding eigenvalues of the reduced
density matrix are $\lambda_1= \cos^2 \alpha$ and
$\lambda_2=\sin^2 \alpha
$,  so Eq. (\ref{HS4}) implies the
 following distribution for the Schmidt angle
\begin{equation}
P( \alpha )=3\cos (2\alpha ) \sin( 4 \alpha).
\label{pchi22}
\end{equation}
For $2 \times 2$ systems ($N=2$) all measures of pure states
entanglement are equivalent and may be expressed
as a function of the Schmidt angle.
For example, the {\sl tangle} reads \cite{Wo98}
  $\tau =4 {\rm det} \rho_A =4\lambda_1\lambda_2=\sin^2(2\alpha)$,
while {\sl concurence} $C$ is equal to $\sqrt{\tau}=sin(2 \alpha)$.
Changing variables in (\ref{pchi22})
we obtain the distributions
\begin{equation}
P( \tau )=\frac{3}{2}\sqrt{1-\tau} \qquad {\rm and} \qquad
P(C)=3C \sqrt{1-C^2} \ .
\label{ptan22}
\end{equation}
consistent with numerical results of Kendon et al. \cite{KNM01}.
First moments of these distributions give the mean
tangle $\langle \tau \rangle _{CP^3}=2/5$
and the mean concurrence
$\langle C \rangle _{CP^3}=3\pi/16\approx 0.589049$,
 averaged in the sense of (\ref{volum3}) over the
natural measure on the space of $N=4$ pure states.

In a similar way it is straightforward to compute the mean
entanglement $E$. It is just the mean von Neumann entropy,
  $H(\varrho)=-{\rm tr } \varrho \ln
\varrho $, of
the reduced density matrices \cite{Ha98}
\begin{equation}
\langle E\rangle _{CP^3}=
\langle E\rangle _{HS}
=\int_{0}^{1/2}[-r\ln r-(1-r)\ln
(1-r)] 24 r^2 dr= \frac{1}{3} .
\label{meanentr}
\end{equation}
In the natural units we use the maximal entanglement
$E_{\rm max}=\ln 2$.
If one expresses the entanglement $E$ in units
of $\ln 2$ then
$E_{\rm max}/\ln 2=1$ and the mean value reads
$\langle E\rangle/\ln 2 =1/(3\ln 2) \approx 0.480898$.
For other orthogonal, unitary and Bures measures we obtain,
 respectively,
$\langle E\rangle_{o}=2\ln2-1 \approx 0.386$,
$\langle E\rangle_{u}=1/2$ and
$\langle E\rangle_{B}=2\ln2-7/6 \approx 0.219$.
In the same
way we compute the mean inverse participation
ratio $R=[{\rm tr}\varrho^{2}]^{-1}$  of the reduced matrices.
 This quantity may vary from
$R=1$ (for pure states) to $R=N$ (for the maximally mixed state), and has
an  interpretation as the
effective number of pure states in the mixture.
For density matrices distributed according to H-S measure
one has $R_{HS}=5/4$.
Performing an analogous integral for the Bures measure we obtain
$R_{B}=8/7$, which reveals high concentration
of this measure on the states of higher purity - see Fig. 1a.
On the other hand, for orthogonal and unitary measures we
find $R_{o}=4/3$ and $R_{u}=3/2$.

\begin{figure}
 \vskip -2.5cm
\hskip 2.0cm
\includegraphics[width=11.5cm,angle=-90]{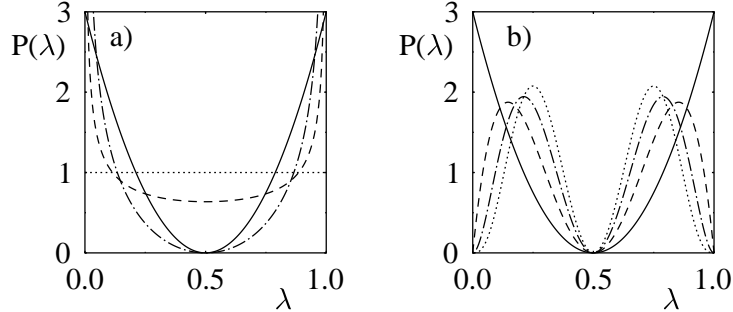}
\vskip -3.0cm
\caption{
Distribution of the eigenvalues of density matrices of size
$N=2$: (a) unitary product measure (uniform distribution -- dotted line),
 orthogonal product
measure  ("cosine distribution" plotted with dashed line), Hilbert-Schmidt
measure (solid line parabola) and Bures  measure (dash-dotted line);
(b) measures $P_{2,K}$ induced by partial tracing;
solid line represents H-S measure i.e. $K=2$,
dashed line $K=3$, dash-dotted line $K=4$, dotted line $K=5$.
}
\label{fig1}
\end{figure}

\section{Measures induced by partial tracing of composite
systems}

\subsection{Joint probability distribution}

Consider a bipartite $N \times K$ composite quantum system.
Pure states
of this system $|\Psi\rangle$ may be represented by a vector
$|\Psi_{j}\rangle=\{\Psi_1,\dots,\Psi_M\}^T$,
where $M=NK$ is the dimension of the
composite Hilbert space
${\cal {H}}={\cal {H}}_A \otimes {\cal {H}}_B$.
In view of the operation of partial tracing
it is convenient to work in a product basis
$|i\rangle \otimes | k\rangle$, where
$|i\rangle \in {\cal H}_A; i=1,\dots,N$ while
$|k\rangle \in {\cal H}_B; k=1,\dots,K$.
A pure state $|\Psi\rangle$ is then represented by a rectangular
complex matrix $\psi_{ik}$.
The normalization condition,
$\|\psi\|^2={\rm tr}\psi\psi^{\dagger}=1$,
is the only constraint imposed on this matrix.
The density matrix $\varrho$, acting on the
composite Hilbert space $\cal H$,
is represented in this basis by a matrix labeled by
four indices,
$\varrho_{ik,i'k'}= \langle i | \otimes \langle k | \varrho
|k'\rangle \otimes |i'\rangle=\psi_{ik}{\psi_{i'k'}}^{*}$.
The partial tracing with respect to the $K$ dimensional subspace ${\cal
H}_B$ gives a reduced density matrix of size $N$
\begin{equation}
\varrho^A={\rm Tr}_B{\varrho} ~ ~ ~ {\rm where} ~ ~ ~
\varrho _{ij}^{A}=\sum_{k=1}^K \psi_{ik} {\psi_{jk}}^{*} \ ,
\label{reducb}
\end{equation}
while a symmetric operation of partial tracing over the
first subsystem leads to a reduced density matrix
$\varrho ^{B}={\rm Tr}_{A}{\varrho }$ of size $K$, where
$\varrho^B_{kl}=\sum_{i=1}^N \psi_{ik} {\psi_{il}}^{*}$.
The natural measure in the space of $M=KN$
dimensional pure states induces then the measure $P_{N,K}(\varrho)$
in the space of
the reduced density matrices (\ref{reducb}) of size $N$.
Note that the problem of measures induced by partial tracing may be
considered as a projection of the
$(KN-1)$ dimensional simplex of eigenvalues ${\cal S}_{KN-1}$
into simplices of smaller dimensions
${\cal S}_{N-1}$ or ${\cal S}_{K-1}$.

Without loss of generality we assume
$K \ge N $, then $ {\varrho}^{A} $ is generically positive definite.
In the opposite case
$ K< N $ the reduced density matrix ${\varrho}^A$ has $N-K$
zero eigenvalues, but the reduced matrix ${\varrho}^B$
of size $K$ is positive definite and has
the same positive eigenvalues.
 In any case we are interested only in
the distribution of the positive eigenvalues.
Let us call the corresponding
positive reduced density matrix again ${\varrho} =\psi
\psi^\dagger$ where  $\psi$ is now considered as a $N \times K$ matrix.
 First we calculate the distribution of matrix elements

\begin{equation}
 P({\varrho}) \propto \int [d\psi ]\  \delta ({\varrho} -\psi
\psi^\dagger)  \ \delta( {\rm Tr}\psi\psi^{\dagger}-1)
\label{Pvonrho}
\end{equation}
where the first delta function is a delta function of
 a Hermitian Matrix and
in the second delta function ${\rm Tr} \psi\psi^{\dagger}$
 may be
substituted by ${\rm Tr}{\varrho}$. Since ${\varrho}$
is positive definite
we can make a
transformation
\begin{equation}
 \psi=\sqrt{{\varrho}}\tilde{\psi},\ \ \ [d\psi]=
{\rm det} {\varrho}^K \
[d\tilde{\psi}]\ .
\end{equation}
Note that $[d\psi]$ includes (alternating) factors $d\psi_{ik} $ and
$d{\psi_{ik}}^*$. The matrix delta function may now be written as
\begin{equation}
\delta(
\sqrt{{\varrho}}(1-\tilde{\psi}\tilde{\psi}^{\dagger})
\sqrt{{\varrho}})\ = \
{\rm det}{\varrho}^{-N} \
\delta(1-{\tilde{\psi}}  {\tilde{\psi}}^{\dagger}) \ ,
\end{equation}
where the first   factor on the right hand side is the inverse Jacobian of
the
 corresponding transformation.
As the result the distribution of matrix
elements is given by
\begin{equation}
P({{\varrho}}) \propto
 \theta({{\varrho}} ) \
\delta( {\rm Tr} {{\varrho}} -1) \ {\rm det} {{\varrho}}^{K-N} ,
\label{Pvonrho1}
\end{equation}
where the theta function assures that ${\varrho}$  is positive definite. It
is then easy to show by the methods of random matrix
theory that the joint
density of eigenvalues $\Lambda=\{ \lambda_1,\lambda_2,\dots,\lambda_N\}$ of
$ {\varrho}$ is given by
\begin{equation}
P_{N,K}(\lambda_1,\dots,\lambda_N)= C_{N,K}
\delta(1-\sum_{i}\lambda_i)\prod_i\lambda_i^{K-N}
\theta(\lambda_i)\prod_{i<j}(\lambda_i-\lambda_j)^2.
\label{P2HS}
\end{equation}
The square of the Vandermonde determinant results from
integrating out the eigenvectors and the normalization
constant ( see also (\ref{constbet})) may be expressed in terms of the Euler
Gamma function $\Gamma(x)$
\begin{equation}
 C_{N,K} =  \frac{\Gamma(KN)}
{\prod_{j=0}^{N-1} \Gamma(K-j) \Gamma(N-j+1) }
\label{constn}
\end{equation}
written here for $K\ge N$.
Observe that for $K=N$ the measure $P_{N,N}$
induced by partial tracing of the pure states in ${\mathbb C}P^{N^2-1}$
coincides with the Hilbert-Schmidt measure  (\ref{HS3}),
so $C_{N,N}=C_N$.
Moreover, it equals to the probability distribution
of squared singular values $\sigma_i^2$ of matrices pertaining to
the
normalized Ginibre
ensemble of complex matrices $A$, where
$\sigma_i$ are given by the square roots
of non-negative eigenvalues of $AA^{\dagger}$.
This is even valid if $A$ is
 a $N \times K$ matrix $(K\ge N)$.
 To clarify this point instead of starting
from equation (\ref{Pvonrho}) we could have been started from a distribution
\begin{equation}
 P({\varrho}) \propto \int [d\psi ]\  \delta ({\varrho} -\psi
\psi^\dagger/{{\rm Tr}\psi\psi^{\dagger}})  \ \exp ( -{\rm Tr}\psi\psi^{\dagger})
\label{Pvonrho2}
\end{equation}
Introducing a delta function for $t={{\rm Tr}\psi\psi^{\dagger}}$
and
rescaling $\psi$ with $\sqrt{t}$ we see that both distributions
(\ref{Pvonrho}) and  (\ref{Pvonrho2}) are equivalent and given by
(\ref{Pvonrho1}).  This follows also from the fact that
$N^2$ complex elements of the random matrix $A$ of the Ginibre ensemble
may be treated
as a random vector distributed according to the natural measure on
${\mathbb C} P^{N^2-1}$ -- see Appendix A.

 In this way
one may give a simple recipe to generate a random density
matrix according to the HS measure and similarly the generalisation
(\ref{Pvonrho1}):

a) prepare a square complex random matrix $A$ of size $N$ pertaining to the
Ginibre ensemble (with real and imaginary parts of
each element being independent normal random variables),

b) compute the matrix $H=AA^{\dagger}/({\rm tr}AA^\dagger )$,
  (generically positive definite).

In principle one could proceed further diagonalizing $H$
and writing the density matrix as $\varrho=UdU^{\dagger}$
where $d$ is the vector of eigenvalues of $H$ and
$U$ is a random unitary rotation drawn according to the Haar measure
$\nu$, so $\varrho$ is taken according to the product measure (\ref{muu}).
However this step is not necessary, since the Ginibre ensemble
is invariant with respect to unitary rotations,
so one may set $\varrho=H$.

The explicit form of the distribution (\ref{P2HS})
for $N=2$ was already derived by Hall \cite{Ha98}.
One could speculate, the larger dimension $K$ of the
auxiliary space ${\cal H}_B$, the more uniform is the induced distribution
$P_{N,K}(\Lambda)$. This is not the case \cite{Ha98},
and the larger $K$ the more is the induced distribution
concentrated in the center of the Bloch ball
- see Fig. 2b.
A similar effect is shown in Fig.2 for $N=3$.
It shows the probability distributions
in the eigenvalues simplex - the equilateral triangle
situated in the plane $\lambda_3=1-\lambda_1-\lambda_2$.
The corners represent three orthogonal pure states,
while the edges denote the matrices of rank $2$.
The more mixed the state, the closer it is to the center of the triangle,
which represents the state $\varrho_*={\mathbb I}/3$.
Unitary product measure, Eq. (\ref{Dirichlet})
with $s=1$, covers uniformly
the entire triangle, while the orthogonal product measure, $s=1/2$,
distinguishes states of higher purity -
see Fig.2a. The Bures measure (\ref{Bures3})
with normalization constant $C'_N=35/\pi$ \cite{Sl99b},
 is even more localized
at states close to pure - see Fig. 2b.
The Hilbert-Schmidt measure (\ref{HS3}), equal to
$P_{3,3}(\Lambda)$, is shown in Fig.2c, while
other
induced distributions (\ref{P2HS}) are plotted in Fig. 2d-2f.
Observe that the larger dimension $K$, the more
the distribution $P_{3,K}(\Lambda)$ shifts
in the eigenvalues simplex toward its center.
Due to the factor $(\lambda_i-\lambda_j)^2$ in (\ref{P2HS})
the degeneracies in spectrum are avoided,
which is reflected by a low probability (white colour)
along all three bisetrices of the triangle.
On the other hand, the distribution
$P_{3,2}(\Lambda)$ is singular,
and located at the sides of the triangle,
which represents the density matrices of rank $2$.
It is equal to $P_{2,3}(\Lambda)$
represented by dashed line in Fig 1b.

\begin{figure}
   \begin{center}
\
 \vskip -0.4cm
\includegraphics[width=10.5cm,angle=0]{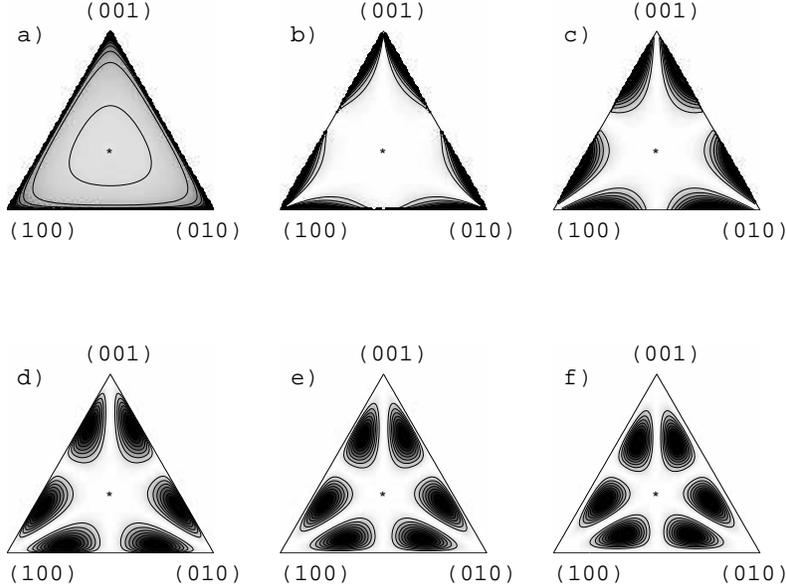}
   \end{center}
\caption{
Distribution of the eigenvalues of density matrices of size
$N=3$ in the simplex of eigenvalues
 (a) orthogonal product measure,
 (b) Bures measure,
 (c) Hilbert-Schmidt measure equal to $P_{3,3}$,
 other measures induced by partial tracing:
  (d) $P_{3,4}$, (e) $P_{3,5}$ and (f) $P_{3,6}$.
}
\label{fig2}
\end{figure}

\subsection{Averages over induced measure}

Let us now calculate some moments
with respect to the Hilbert-Schmidt measure
$P_{HS}(\Lambda)=P_{N,N}(\Lambda)$.
 Taking Fourier representation  of the
 delta function $\delta(1-\sum_i\lambda_i)$ one
sees that averages which are homogeneous functions of some ${\lambda}_k$ are
related to the corresponding moments of the Laguerre ensemble:
\begin{equation}
P_L(\lambda_1,...\lambda_N)\propto
\exp(-\sum_{i}\lambda_i)\prod_i\lambda_i^{K-N}
\theta(\lambda_i)\prod_{i<j}(\lambda_i-\lambda_j)^2.
\end{equation}
Here simple moments can be calculated by the method of orthogonal
polynomials \cite{Mehta}. For example one finds for $K=N$ and $\nu>-1$
\begin{equation}
\langle {\rm Tr}{\varrho}^{\nu}\rangle_{HS} \ =\ B_{N,\nu}
 \int_0^{\infty}{\rm d} x \
 x^{\nu}\sum_{n=0}^{N-1} L_n(x)^2 {\rm e}^{-x} \ ,
\label{rhonu}
 \end{equation}
with the Laguerre polynomials defined by
\begin{equation}
 L_n(x)= {\rm e}^x \frac{1}{ n!}
(\frac{d}{ d x} )^n x^n {\rm e}^{-x}
\end{equation}
and $B_{N,\nu}$ a constant which results from the
 map of the Hilbert Schmidt measure
to the Laguerre measure:
\begin{equation}
B_{N,\nu}=\Gamma(N^2)/\Gamma(N^2+\nu).
\end{equation}
Note that we also obtain an explicit expression for the entropy
$H(\varrho)=-\langle{\rm
Tr}{\varrho}{\rm ln}{\varrho}\rangle$ by taking the derivative
of Eq.(\ref{rhonu}) at $\nu=1$. This can be
 written as a sum of Digamma  functions.
For large $N$ it displays asymptotic behaviour
\begin{equation}
\langle H({\varrho})\rangle_{HS} \approx
\ln N - \frac{1}{2} +O\bigl(\frac{\ln N}{N} \bigr).
\label{entrasym}
 \end{equation}
This formula gives one of the main results of this paper --
the mean entanglement of pure states of the $N \times N$
bipartite system, averaged over the natural measure
on the space of pure states.
It is equivalent to
$\langle H({\varrho})\rangle_{HS} \approx \ln(0.6065N)$, which
stays in agreement with recent numerical results of
Lakshminarayan \cite{La00}, who computed mean entanglement of pure
states of an exemplary chaotic quantum system -- the four dimensional,
generalized standard map without time reversal symmetry.
 Our result is somewhat
similar to the asymptotics of the mean
entropy of complex random vectors
$\langle H(|\phi\rangle )\rangle_{{\mathbb C}P^{N-1}} \approx
\ln N - 1+\gamma$ \cite{Jo90},
where the Euler constant $\gamma\sim 0.5772$.

The recurrence relation
\begin{equation}
 xL_n(x)=-(n+1)L_{n+1}(x) + (1+2n)L_n(x)-nL_{n-1}(x)
\end{equation}
implies for $\nu=2$
\begin{equation}
\langle {\rm Tr}{\varrho}^2\rangle_{HS}
=\frac{1}{N^2(N^2+1)}\sum_{n=0}^{N-1}((n+1)^2+(1+2n)^2+n^2)
\end{equation}
Finally the sum over $n$ can be done with the result:
\begin{equation}
\langle{\rm Tr}{\varrho}^2\rangle_{HS}=\frac{2N}{ N^2+1}
\end{equation}
Using the recurrence relation for the $L_n(x)$ one can find the term
$\propto L_n(x)$ in the Laguerre
expansion of $x^{\nu}L_n(x)$ for
large $n$. From this we find the following
asymptotic behavior for large $N$:

\begin{equation}
\langle {\rm Tr} {\varrho}^{\nu}\rangle_{HS}
 =   \frac{\Gamma(1+2\nu)\  N^{1-\nu}}{\Gamma(1+\nu) \Gamma(2+\nu)}\
(1+O(\frac{1}{ N})).
\end{equation}
This implies for the
entropy an asymptotic behavior (\ref{entrasym}).
Inspecting the recurrence relation further one sees that the moments in
question are always given by the ratio of two polynomials of $N$. With
this knowledge it is easy to calculate some further moments:
\begin{equation}
\langle {\rm Tr}{\varrho}^3 \rangle_{HS} \ =\
\frac{5N^2+1}{(N^2+1)(N^2+2)}
\end{equation}
and
\begin{equation}
\langle {\rm Tr}{\varrho}^4\rangle_{HS} \ =\
\frac{14N^3+10N}{(N^2+1)(N^2+2)(N^2+3)}.
\end{equation}
Finally let us mention that the result for the second moment can be
generalised for density matrices distributed according to (\ref{P2HS})
for arbitrary $N$ and $K$
\begin{equation}
{\langle {\rm Tr}{\varrho}^2\rangle}_{N,K} \ =\ {\frac{N+K}
{NK+1}} \ .
\label{rhokn}
\end{equation}
The calculation leads to the generalised
Laguerre polynomials. Due to the symmetry in $K$ and $N$
it is even not necessary to mention what
is the largest of both dimensions $K$ or $N$.
The participation ratio reads thus
$R=(NK+1)/(N+K)$ and is consistent with recent
results of Zanardi et al. \cite{ZZF00}. They computed the
linear entropy $1-{\rm tr} \varrho^2$
for an ensemble of mixed states obtained
by partial tracing of pure states
$U(|\psi_1\rangle \otimes |\psi_2\rangle)$,
where $|\psi_1\rangle$ and $|\psi_2\rangle$
are pure states distributed according to
the natural measure on
${\mathbb C}P^{N-1}$ (${\mathbb C}P^{K-1})$,
and $U$ is a random unitary matrix generated according to
the Haar measure on $U(NK)$. It is easy to see that this
construction leads to the induced measure $P_{N,K}$
given by (\ref{P2HS}), since
the Haar measure on $U(NK)$ is invariant with respect
to multiplication by product matrices
$U(N)\otimes U(K)$ representing local operations.
Thus our results concern not only average entanglement
of pure states, but in spirit of
\cite{ZZF00} also averaged {\sl
entangling power} of global unitary operators.

\section{Generalized ensembles of density matrices}

One may obtain a different ensemble of density matrices starting from a real
pure state with time reversal symmetry
 \begin{equation}
{{ \varrho}}
 = \sum_{k=1}^K u_{ik} {u_{jk}}
\end{equation}
  with real $u_{ik}$. The only constraint we have now is
$ \sum_{ik}u_{ik}^2
 = 1 $.
Again we may restrict to $K\ge N$ and since ${\varrho}$ is
positive definite we  can make the transformation
\begin{equation}
u=\sqrt{{\varrho}} {\tilde u}, ~ ~ ~
\end{equation}
Now the matrix-delta function of
 the real symmetric matrix transforms according to
\begin{equation}
\delta(\sqrt{{\varrho}}
(1-\tilde{u}\tilde{u}^{\top})\sqrt{{\varrho}})\ =\
{\rm det}{\varrho}^{(-N-1)/2}\ \delta(1-\tilde{u}\tilde{u}^{\top})\ ,
\end{equation}
such that
\begin{equation}
 P({\varrho}) \propto
\theta({\varrho})
\ \delta({\rm Tr} {\varrho}-1)
\ {\rm det}{\varrho}^{(K-N-1)/2}
\end{equation}
and the distribution of eigenvalues is given by
\begin{equation}
P^{(1)}(\lambda_1,\dots,\lambda_N)\propto\delta(1-\sum_{i}\lambda_i)
\prod_i\lambda_i^{(K-N-1)/2}\theta(\lambda_i)\prod_{i<j}|\lambda_i-\lambda_j|.
\end{equation}
Thus $P_{N,K}=P^{(2)}_{N,K}$ corresponds to unitary symmetry with $\beta=2$
and $P^{(1)}$ to
 orthogonal symmetry with $\beta=1$. Both are special cases of

\begin{equation}
P^{(\beta)}_{N,K}\ = \
C^{(\beta)}_{N,K}
\delta(1-\sum_{i}
\lambda_i)\prod_i\lambda_i^{(\beta(K-N)+\beta-2)/2}
\theta(\lambda_i)\prod_{i<j}|\lambda_i-\lambda_j|^{\beta}
\end{equation}
which can also be given a meaning in the case of symplectic symmetry
$\beta=4$ (in that case each Kramers degenerate eigenvalue is counted once
and rescaled).
In the general case of arbitrary real $\beta>0$ the normalization constant
reads \begin{equation}
 C^{(\beta)}_{N,K} =  \frac{\Gamma(KN\beta/2) (\Gamma(1+\beta/2))^N}
{\prod_{j=0}^{N-1} \Gamma((K-j)\beta/2) \Gamma(1+(N-j)\beta/2)} \ .
\label{constbet}
\end{equation}
It can be derived using the map to the
 Laguerre ensemble and using Selberg's
integral \cite{Mehta}.
These ensembles of density matrices
may be constructed by generating
real ($\beta=1$) or  complex ($\beta=2$)
random Gaussian matrices $A_{\beta}$ and
defining random density matrices by
$\varrho=A_{\beta}A_{\beta}^{\dagger}/({\rm
tr}A_{\beta}A_{\beta}^\dagger )$. In the case of symplectic symmetry
$(\beta=4)$ one has complex matrices $A_{\beta}$ but then takes only the
selfdual part of $\rho$. In fact, any density $P(A)$,
in the space of complex density matrices of size $N$
generates in this way a certain ensemble of density matrices
 of this size.

Let us mention that non-negative matrices $AA^{\dagger}$
are sometimes called random {\sl Wishart matrices}.
The joint probability distribution of their
eigenvalues (i.e. the distribution of singular values of $A$)
was analyzed in \cite{BFP98,SM99}.
Real random
matrices $A_1A_1^{T}$
may describe the matrix of correlations between
time series of different stock prices in the presence of noise.
This fact explains a recent interest in their spectral properties
(or in the distribution of singular values of
real random matrices $A_{1}$),  from the point of view
of mathematical finances \cite{LCBP99,PGRAS99,NJ}.
Thus the theory of random matrices
provides an unexpected link between
the computation of the mean entanglement
of a certain ensemble of composite quantum states
and the estimation of the financial risk \cite{Bou}.

\section{Concluding remarks}

We described various ways of defining ensembles in the space of density
matrices of finite size $N$. While it is natural to take eigenvectors
distributed according to the Haar measure on $U(N)$, there
are several possibilities of introducing the measure into the $(N-1)$
dimensional simplex of eigenvalues.
Besides measures related to random vectors
(of orthogonal or unitary matrices), one may
define measures related to distances in the space of
density matrices (say, Hilbert-Schmidt measure or Bures measure).

An alternative way consists in
assuming certain measures in more dimensional spaces
(e.g. in the space of pure states in $NK$ dimensions
or in the $N^2$ dimensional space of square matrices of size $N$),
and considering the measures induced by projection
onto the simplex.
We generalized the results of Hall \cite{Ha98}
deriving an explicit formula for $P_{N,K}$
and proved that the measure induced by symmetric partial tracing,
$P_{N,N}$ coincides with the Hilbert-Schmidt measure (\ref{HS3}).
We succeeded in computing some averages with respect to this measure
including an explicit formula for the Shannon entropy of the reduced
density matrix, equal to the mean entropy of entanglement,
 averaged over the manifold of pure states of the composite
$N \times N$  system.

Moreover, we demonstrated a link between the H-S measure
determining the spectra of the reduced density matrices
and the
singular values of the random matrices distributed according to the
Ginibre ensemble. As a by--product we demonstrated
that $N^2$ random complex (real) numbers drawn with respect to the
normal distribution, which constitute a non-Hermitian (non-symmetric)
random matrix,  can be considered as a vector
of the unitary (orthogonal) random matrix of size $N^2$,
drawn according to the Haar measure on $U(N^2)$ (on $O(N^2)$).
It would be interesting to find an ensemble of random matrices
corresponding to Bures measure (\ref{Bures3}),
to compute the normalization constants $C_N'$ for arbitrary $N$
and to evaluate the mean entanglement averaged with respect to this
measure.

It is a pleasure to thank Pawe{\l}~Horodecki, Marek Ku{\'s}
and Thomas Wellens for inspiring discussions.
Financial support by Komitet Bada{\'n} Naukowych in Warsaw under
the grant 2P03B-072~19,  the Sonderforschungsbereich 'Unordnung
und gro{\ss}e Fluktuationen'  der Deutschen Forschungsgemeinschaft
and the European Science Foundation is gratefully acknowledged.

Note added: After this work was completed we learned form M.J.W. Hall
about related papers published on this subject:
the distribution (\ref{P2HS}) was derived by Lloyd and Pagels \cite{LP88},
while the formula for the mean entropy averaged over
the induced measures was
conjectured by Page \cite{Pa} and proved by Sen \cite{Se96},
(see note added to \cite{Ha98}, preprint ArXiv quant-ph/9802052).
We are grateful to M.J.W. Hall for helpful correspondence.

\appendix

\section{Rescaling of random Gaussian variables}

We are going to analyze the joint probability distribution
of independent random numbers $y'_i$, rescaled according to
\begin{equation}
y'_i \rightarrow
y_i=\frac{y'_i}{\sum_{i=1}^N y'_i}\ .
\label{resc}
\end{equation}
We assume that each random number is drawn according to the same
probability distribution $P(y')$ defined for positive $y'$.
It is easy to see that the joint distribution of the rescaled variables,
$P(y_1,...,y_N)$, defined in the
interval $I=[0,1]$, is symmetric with respect to
exchange of any variables, $y_i \longleftrightarrow  y_k$.
It is worth to point out that the {\sl uniform}
initial distribution, $P(y'_i)=1$ for $y'\in[0,1]$
does not produce the uniform distribution of the rescaled
variables, but
$P(y_1....y_N) = \delta(1-\sum_iy_i)/(N  ({\rm max}\{y_i\})^N)$.
In the simplest case, $N=2$, it gives
$P(y)=1/(2(1-y)^2)$ for $y <1/2$, (where $y=y_1=1-y_2$)
and symmetrically for $y>1/2$.

Let us now assume that the random numbers are given by
 squares of Gaussian random numbers,
$y_i'=x_i^2$ and $P(x_1,x_2,...x_N)\propto \exp(-\sum_i x_i^2)$.
We want to calculate the distribution of the rescaled variables $ y_i=
x_i^2/\sum_ix_i^2$:
\begin{equation}
P(y_1,y_2,...y_N)\propto \int {\rm d}x_1...{\rm d}x_N\exp(-\sum_ix_i^2)
\prod_i\delta(y_i-x_i^2/\sum_ix_i^2) \ .
\end{equation}
Introducing a $\delta$ function for the constraint
$t= \sum_ix_i^2$ we have
\begin{equation}
P(y_1,y_2,...y_N)\propto \int_0^{\infty}{\rm d}t{\rm e}^{-t} \int {\rm
d}x_1...{\rm d}x_N \delta(t-\sum_ix_i^2)\prod_i\delta(y_i-x_i^2/t)
\end{equation}
If we now rescale $x_i\to x_i\sqrt{t}$ we obtain
\begin{equation}
P(y_1,y_2,...y_N)\propto \int_0^{\infty}{\rm d}t{\rm e}^{-t}
t^{-1+N/2}\int {\rm d}x_1...{\rm d}x_N
\delta(1-\sum_ix_i^2)\prod_i\delta(y_i-x_i^2)
\end{equation}
which yields the Dirichlet distribution
(\ref{Dirichlet}) with $s=1/2$.

This result may be generalized for
the
sum of $\beta$ squared random numbers
drawn according to the normal distribution,
$y_i' = \sum_{j=1}^{\beta}x_{i,j}^2$,
the distribution of the sum is given by the $\chi^2_{\beta}$
distribution with $\beta$ degrees of freedom.
In an analogous way we obtain that the joint distribution of
 $y_i$'s rescaled as in (\ref{resc})
is given by the Dirichlet distribution
\begin{equation}
P(y_1,y_2,..., y_N) \propto  \delta(1-\sum_i y_i)
\prod_{i=1}^N
 y_i^{\beta/2-1},
\end{equation}
with $s=\beta/2$.
The most important case, $\beta=2$,
shows that the distribution of squared components
of a rescaled complex vector of Gaussian random numbers
is {\sl uniform} in the $(N-1)$ dimensional simplex.

Above results find following simple applications.
Real (complex) random
vectors, i.e. columns (or rows) of orthogonal
(unitary) random matrix may be constructed out of independent random Gaussian
numbers (real or complex, respectively) rescaled
 according to Eq. (\ref{resc}).
 This is
consistent with the well known fact that the distribution of squared
components of eigenvectors of a GOE (GUE) matrix are given by the
$\chi_{\beta}^{2}$ distribution with $\beta=1$
(or $2$) degrees of freedom \cite{Mehta}. Thus certain
averages, one wants to compute averaging over the set of pure states (say,
entropy, participation ratio, etc),  may be replaced by averages over the
simplex in $R^{N-1}$ with the Dirichlet measure (\ref{Dirichlet})
with $s=1/2$ or $s=1$.

Furthermore, a complex square random matrix $\psi_{kl}^{\prime}$ of
size $N$ pertaining to the Ginibre ensemble
after rescaling
 $\psi^{\prime}\rightarrow\psi=
\psi^{\prime}/\sqrt{{\rm tr}{\psi^{\prime}}^{\dagger}\psi^{\prime}}$
represents a random pure state drawn according to the
natural measure on ${\mathbb C}P^{N^2-1}$.
A similar link is established between
rescaled rectangular matrices of size $K \times N$
and the elements of ${\mathbb C}P^{KN-1}$.
Analogously, normalized non-symmetric real random matrices with
each element drawn independently
according to the normal distribution
are equivalent to random vectors distributed with respect to
the natural measure on the
real projective space, ${\mathbb R}P^{N^2-1}$.

\end{document}